\documentclass{ws-ijmpa}

\usepackage{amssymb}
\usepackage{amscd}
\begin{document}


\markboth{G.Marmo, A.Simoni and F.Ventriglia} {Quantum Systems and
Alternative Unitary Descriptions}

%
\catchline{}{}{}{}{}
%

\title{QUANTUM SYSTEMS AND \\ ALTERNATIVE UNITARY DESCRIPTIONS }

\author{\footnotesize G. MARMO}

\address{Dip. Scienze Fisiche and INFN Sez. di Napoli, Universit\`a Federico II ,
 Compl. Univ. Monte S.Angelo\\
Napoli, 80126, Italy \footnote{marmo@na.infn.it} }

\author{A. SIMONI}

\address{Dip. Scienze Fisiche and INFN Sez. di Napoli, Universit\`a Federico II ,
 Compl. Univ. Monte S.Angelo\\
Napoli, 80126, Italy \footnote{simoni@na.infn.it}}

\author{F. VENTRIGLIA}

\address{Dip. Scienze Fisiche and INFM Unit\`a di Napoli, Universit\`a
Federico II , Compl. Univ. Monte S.Angelo\\
Napoli, 80126, Italy \footnote{ventriglia@na.infn.it} }

\maketitle

\pub{Received (Day Month Year)}{Revised (Day Month Year)}

\begin{abstract}
Motivated by the existence of bi-Hamiltonian classical systems and
the correspondence principle, in this paper we analyze the problem
of finding
 Hermitian scalar  products  which turn  a given flow on
a Hilbert space into a unitary one. We show how different
invariant Hermitian scalar products give rise to different
descriptions of a quantum system  in the Ehrenfest and
 Heisenberg picture.

\end{abstract}

\section{\protect\bigskip Introduction}

Classical bi-Hamiltonian systems have played a relevant role in
the past decades for the study of completely integrable systems,
both for finite and infinite number of degrees of freedom (see
Ref. 1 and Ref. 2 for quantum systems).

Quantum systems admitting alternative commutation relations have
been considered many times since the pioneering paper of E.P.
Wigner\cite{Wigner}, see also Ref. 4, 5, 6.

Quantum systems described by non Hermitian operators and
possessing a real spectrum have been analyzed by several authors
(e.g. Ref. 7, 8).

In a recent paper\cite {Morandi} we have shown that these
situations may be better tackled within the framework of the
''inverse problem for quantum systems''. From a mathematical point
of view a similar problem was first discussed  by Nagy\cite{Nagy}
long ago.

To clearly formulate this problem let us briefly recall first the
symplectic inverse problem for classical linear systems. Starting
with a vector field
\begin{equation}
\Gamma =\Gamma ^{i}\frac{\partial }{\partial \xi ^{i}}\ \ \ ,
\end{equation}
one searches far all possible Hamiltonian descriptions in terms of
symplectic structures
\begin{equation}
\omega =\omega _{jk}d\xi ^{j}\wedge d\xi ^{k}\ \ \ \ \ \ ,
\end{equation}
by solving for the equation $L_{\Gamma }\omega =0$ . Every
symplectic structure admits an ''inverse'', it is a bivector
field, usually called a Poisson tensor, defined by $\Lambda
^{ik}\omega _{kj}=\delta _{j}^{i}.$
Thus, the inverse problem amounts to search for all decompositions of $%
\Gamma $ as the product
\begin{equation}
\Gamma ^{i}=\Lambda ^{ik}\frac{\partial H}{\partial \xi ^{k}}\ \ \ \ .
\end{equation}
Where $\Lambda ^{ik}$ is a skewsymmetric real, point dependent, matrix. If
we deal with linear vector fields and quadratic Hamiltonians, say
\begin{equation}
\Gamma ^{i}=A_{k}^{i}\xi ^{k}\ \ \ ,\ \ \ \ H=\frac{1}{2}\xi ^{k}H_{kj}\xi
^{j}\ \ ,
\end{equation}
the inverse problem becomes a problem of linear algebra, i.e.
searching for all decompositions of the dynamical matrix $A$ into
the product of a non degenerate skew-symmetric matrix $\Lambda $
and a symmetric matrix $H$, in compact form $A=\Lambda \cdot H.$ \
\ When $\Lambda $ is not required to be non-degenerate, we are
dealing with the ''inverse problem for Poisson dynamic''.\cite{??}
 We should remark \ that this problem is more interesting when we
are considering non-linear situations and non constant rank of
$\Lambda $, otherwise, by quotienting with respect to the kernel
of $\Lambda ,$ we may go back to the symplectic situation.

Thus all possible decompositions of $A$, in the stated form, provide us with
alternative Hamiltonian descriptions. In many physical situations, when
dealing with completely integrable systems, we are interested in the
existence of alternative decompositions once one has been already given,
i.e. we search for alternative Hamiltonian descriptions for a given
Hamiltonian system $\Gamma $. The alternative descriptions would
characterize $\Gamma $ as a multi-Hamiltonian vector field. It is almost
obvious that many alternative Hamiltonian descriptions will be generated by
symmetries for $\Gamma $ which are not canonical transformations for $%
\Lambda $. However this way of generating new Hamiltonian
descriptions will not exhaust the class of alternative ones. For
instance\cite{Ventriglia} the two dimensional isotropic Harmonic
Oscillator has different decompositions with either $H$ definite
positive or with signature (+,+,-,-), which arise from the
following Hamiltonian descriptions:
\begin{equation}
\begin{array}{l}
H_{0}=\frac{1}{2}(p_{1}^{2}+q_{1}^{2}+p_{2}^{2}+q_{2}^{2}) \\
\omega _{0}=dp_{1}\wedge dq_{1}+dp_{2}\wedge dq_{2}
\end{array}
\end{equation}
and
\begin{equation}
\begin{array}{l}
H=\frac{1}{2}(p_{1}^{2}+q_{1}^{2}-p_{2}^{2}-q_{2}^{2}) \\
\omega =dp_{1}\wedge dq_{1}-dp_{2}\wedge dq_{2}\ .
\end{array}
\label{4}
\end{equation}
These factorizations cannot be related by any similarity
transformation. In this setting, it has been shown\cite{Giordano}
that when the critical point of the linear vector field $\Gamma $
is stable in the Liapunov sense, among the alternative Hamiltonian
descriptions there is one which is positive definite.

The formulation of the inverse problem for quantum systems is now quite
natural: given a vector field $\Gamma $ on some vector space $V$, we search
for all Hermitian structures $h$ which are solution of the equation $%
L_{\Gamma }$ $h=0$ .

In our previous papers Ref. 7, 17, to avoid technicalities, we
have carried on our study of the inverse problem within the
framework of finite dimensional Hilbert spaces; in this paper we
would like to consider the problem in the more relevant case of
infinite dimensional Hilbert space.

This paper is organized as follows: in Sections 2, 3, 4, 5 the
problem is specified, mathematical results are reviewed, some
applications are given and the dependence of results on the
fiducial metric $\ h_{0}$ \ is studied. In Section 6 we give some
results on the existence of alternative invariant Hermitian scalar
products; in Section 7 a way for finding invariant scalar products
for Abelian groups is considered; in Section 8 we deal with the
more general situation of related operators instead of similar
ones; in Section 9 we apply our considerations to the Heisenberg
group. Finally in Section 10 we describe quantum systems as
Hamiltonian systems and discuss the consequences of the existence
of alternative invariant scalar products, in Section 11 we draw
some conclusions. Results which are available in the literature
are referred here as Theorems, while our considerations are recast
in Propositions and Corollaries.

\section{The Inverse Problem for Quantum Systems}

We consider a linear dynamical system $\Gamma $ on a complex vector space of
states $\mathbb{H}$ carrying a Hilbert space structure $h_{0}$ \ (we define
all scalar products to be linear in their second factor). Then $\Gamma $%
\begin{equation}
\Gamma :\mathbb{H}\rightarrow T\mathbb{H\ =}\ \mathbb{H\times H\ \ },\mathbb{%
\ \ }\Gamma (\Psi )=\mathbb{(}\Psi ,-\frac{i}{\hbar }H\Psi )
\end{equation}
determines the differential equation
\begin{equation}
\frac{d}{dt}\psi =-\frac{i}{\hbar }H\psi \ \ ,  \label{Scroe}
\end{equation}
where $H$ is a linear operator on $\mathbb{H}$ ,\ but we do not require any
Hermiticity properties with respect to $h_{0}$. Hereafter we put $\hbar =1.$

The inverse problem consists of finding which conditions have to be
satisfied by $\Gamma $ for the existence of Hermitian scalar products
invariant under the time evolution associated with $\Gamma $ .

In finite dimensions, i.e. for a complex linear vector field on $\mathbb{C}%
^{n}$ , we have\cite{Ventriglia} the following Proposition:

\begin{proposition} A complex linear vector field $\Gamma (\Psi )=\mathbb{(}\Psi ,-%
iH\Psi )$ generates a flow $\phi _{t}:$ $\mathbb{C}%
^{n}\rightarrow \mathbb{C}^{n}$ preserving some Hermitian scalar product $h$%
, i.e. $\phi _{t}^{\ast }h=h$, iff any one of the following
equivalent conditions is satisfied:

1) $H=H^{\dagger },$ where the adjoint is taken with respect to
the scalar product defined by $h$ , i.e. $L_{\Gamma }h=0$;

2) $H$ is diagonalizable and has a real spectrum;

3) all the orbits $e^{-iHt}\Psi $ are bounded sets, for any
initial condition $\Psi $.
\end{proposition}

\textbf{Remark. } Sometimes, properties stated in 2) are derived
by requiring that the Hamiltonian is $\mathcal{PT}-$symmetric (see
for instance Ref. 12, 13).

The proof of this Proposition relies on the existence of a
decomposition of $H$ into a nilpotent part and a semisimple part,
commuting among themselves. Once we exponentiate this
decomposition, the boundedness condition rules out the nilpotent
part and requires that the semisimple part of $-iH$ has only
imaginary eigenvalues. When going to infinite dimensions one may
try to use a similar procedure, however the corresponding
separation of $-iH$ holds true only for a special class of
operators.\cite{Dunford} Therefore we have to use a different
approach.

When dealing with infinite dimensions, it is clear that, according
to Weyl,\cite{Weyl} condition 3) will play a more convenient role
because it is stated in terms of (what is going to become) the
bounded operators $e^{-iHt}$ instead of the infinitesimal
generator $-iH$ which, in the most physical situations, turns out
to be unbounded and therefore would raise domain problems which
make statements more cumbersome.

Therefore, within Weyl's ideology, it is better to deal with
finite transformations (i.e. automorphisms of the state space)
rather than infinitesimal transformations (i.e. endomorphisms,
which, in general, create domain problems). This step, when
starting with Eq. (\ref{Scroe}), may be achieved by using the
Cayley map, i.e. by replacing $H$ with
\begin{equation}
T=(H-i\mathbb{I)}(H+i\mathbb{I)}^{-1}\ .  \label{Cayley}
\end{equation}
For a recent, authoritative analysis of this map see
Kostant-Michor.\cite {Michor} In this way we search for scalar
products turning $T$ into a unitary operator and accordingly $H$
into a self-adjoint operator with a unitary flow $e^{-iHt}$.

Of course we could also decide to formulate the inverse problem in quantum
mechanics directly in terms of one-parameter groups of automorphisms for the
state space and to seek for all Hermitian products which turn the
one-parameter group of transformations into a unitary one-parameter group.
We shall therefore start with automorphisms of the state space instead of
endomorphisms. To set the stage, we consider the vector space of states to
be a Hilbert space, i.e. topology, completeness and bases are defined with
respect to a chosen fiducial scalar product $h_{0}$. However this Hermitian
structure need not be invariant under the transformations we are dealing
with.

In the next Section we review few relevant results scattered in
the existing literature, they were motivated by the search for
stability criteria for infinite dimensional systems\cite{Krein}
(compare the paragraph after formula Eq. (\ref{4}) of the
Introduction).

\section{Uniformly Bounded Operators}

Inspired by condition 3) of the above Proposition 1 we may
consider an automorphism $T$ \ of a Hilbert space $\mathbb{H}$
with a Hermitian scalar product $h_{0}$ and construct the orbits
\begin{equation}
\left\{ T^{k}\Psi \right\} \ \ \ \ ;\ \ \ k\in \{0.\pm 1,\pm 2,\dots \}
\end{equation}
and require that all of them, with respect to the norm induced by
$h_{0}$ , are bounded sets for any value of $\Psi $. The use of
the principle of uniform boundedness\cite{Simon} shows that this
is equivalent to require that $T$ is uniformly bounded.

We recall that the automorphism $T$ on $\mathbb{H}$ is said to be uniformly
bounded if there exists an upper bound $c<\infty $ such that
\begin{equation}
||T^{k}||<c\ \ ;\ \ \ k\in \{0.\pm 1,\pm 2,\dots \}\ \ .  \label{unifbound}
\end{equation}
Condition (\ref{unifbound}) is called Nagy criterion. For such an operator $%
T $ the following Theorem\cite{Nagy} holds:

\begin{theorem} For a uniformly bounded operator $T$ \ there
exists a bounded positive selfadjoint transformation $Q$ such that
\begin{equation}
\frac{1}{c}\mathbb{I}\leq Q\leq c\mathbb{I}
\end{equation}
and $QTQ^{-1}=U$ \ is unitary with respect to the fiducial $h_{0}$. This
implies that $T=$ $Q^{-1}UQ$ is unitary with respect to
\begin{equation}
h_{T}(X,Y):=h_{0}(Q^{2}X,Y)\ .
\end{equation}
\end{theorem}

\begin{proof}(Sketch) The essential idea of the proof is to
define the invariant scalar product $h_{T}(X,Y)$ as the limit, for
$n$ going to infinity, of $\ h_{0}(T^{n}X,T^{n}Y)=:h_{n}(X,Y).$
This is the limit of a bounded sequence of complex numbers which
does not exist in general, at least in the usual sense. Therefore
a generalized concept of limit for bounded sequence, introduced by
Banach and Mazur\cite{Banach}, has to be used. This generalized
limit (denoted as $Lim$) amounts to define the invariant scalar
product $h_{T}$ as the transformed scalar product $h_{n}$
''at infinity'' , where $T$ is interpreted as the generator of a $\mathbb{Z}%
- $action on $\mathbb{H}$.
\end{proof}
It is possible\cite{Nagy} to use the same approach to deal with an $\mathbb{%
R}-$action instead of the $\mathbb{Z}-$action so that:

\begin{theorem} When the one-parameter group of automorphisms
$T(s)$ of linear transformations is uniformly bounded, that is
$||T(s)||<c ,\ \  s\in (-\infty ,\infty )\ ,$ there exists a
bounded selfadjoint transformation $Q$ such that
$QT(s)Q^{-1}=U(s)$ is a one-parameter group of \ unitary
transformations. Clearly continuity properties with respect to $s$
are the same for both $T(s)$ and $U(s)$.
\end{theorem}

\textbf{Remark.}The iterated application of the uniformly bounded operator $%
T $ may be viewed as a discrete time evolution, which, according
to Theorem 2, can be made unitary. Here we note that such a
discrete time evolution may be fitted within a continuous
differentiable time evolution. We have the following

\begin{proposition}If $T$ is a uniformly bounded operator then there
exists a
bounded operator $A$, selfadjoint with respect to the invariant product $%
h_{T},$ such that $T^{n}=e^{iAn}$ for any $n\in \mathbb{Z}$.
\end{proposition}

\begin{proof} From Theorem 2 we know that $T$ is $h_{T}-$unitary
so that it can be written as

\begin{equation}
T=\int\limits_{0}^{2\pi }e^{i\lambda }dE_{\lambda }^{T}\ \ \ ,
\end{equation}
where $E_{\lambda }^{T}$ is the uniquely defined spectral family
for $T$ with the property $E_{0}^{T}=0$ and $E_{2\pi
}^{T}=\mathbb{I}$.

Now define $A=\int\limits_{0}^{2\pi }\lambda dE_{\lambda }^{T}$ \ , so that $%
T^{n}=e^{iAn}$ . $\ $The operator $A$ is of course
$h_{T}-$selfadjoint, bounded and defined on the entire Hilbert
space.
\end{proof}

Of course, the one-dimensional unitary group $e^{iAt}$ is a
continuous one-parameter group containing the discrete subgroup
$\left\{ T^{n}\right\} $ and moreover all orbits $e^{iAt}\Psi $ ,
for any $\Psi \in \mathbb{H}$ , are differentiable in $t$ and
solve\ the Schroedinger equation for $A.$ It is possible to go
from one-parameter groups of transformations to
finitely-many-parameters groups of transformations if they define
an Abelian group. The following Theorem (see for instance Ref. 20)
holds:

\begin{theorem} A uniformly bounded action on $\mathbb{H}$ of
an Abelian group $\mathcal{G}$ , i.e.
\begin{equation}
||G||<c\ \ \ \ ,\ \ \ \ \forall \ G\in \mathcal{G}
\end{equation}
can be turned into a unitary action with the help of a bounded selfadjoint
transformation $Q,$ i.e. $QGQ^{-1}$ are unitary transformations for any $\
G\in \mathcal{G}$.
\end{theorem}

The proof of this Theorem cannot use the idea of the generalized
Banach limit, like in the previous more restricted cases, because
no assumptions are made on the structure and the topology of
$\mathcal{G}$ . In fact, the proof relies on a very general result
on the existence of fixed points for any continuous transformation
of convex sets (Ref. 21, 22, see also Ref. 23).

For practical purposes, when Nagy condition is not easy to check,
some equivalent conditions may be used (see Ref. 24, 25). It has
been shown that Nagy condition Eq. (\ref{unifbound}) is equivalent
to
\begin{equation}
\begin{array}{l}
\sup\limits_{|\lambda |>1}(|\lambda |^{2}-1)\int\limits_{0}^{2\pi
}||(T-\lambda )^{-1}u||^{2}d\theta \leq C||u||^{2} \\
\sup\limits_{|\lambda |>1}(1-|\lambda |^{-2})\int\limits_{0}^{2\pi
}||(T^{\dagger }-\lambda ^{-1})^{-1}u||^{2}d\theta \leq C||u||^{2}
\end{array}
\end{equation}
where $\theta = \arg \lambda$. By using the Cayley transform a
condition for the similarity of an operator $L$ to a selfadjoint
one is recovered as:
\begin{equation}
\begin{array}{l}
\sup\limits_{\varepsilon >0}\varepsilon \int\limits_{R}^{{}}||(L-\lambda
)^{-1}u||^{2}dk\leq C||u||^{2} \\
\sup\limits_{\varepsilon >0}\varepsilon \int\limits_{R}^{{}}||(L^{\dagger
}-\lambda )^{-1}u||^{2}dk\leq C||u||^{2}
\end{array}
\end{equation}
where $\lambda =k+i\varepsilon $.

In Theorems 2, 3, 5 the commutativity hypothesis is
crucial.\cite{Pisner} In the non commutative case one needs
necessarily some other assumptions, for instance that
$\mathcal{G}$ is a (representation of a) compact group. Then the
existence of the invariant Haar measure guarantees that
$\mathcal{G}$ is similar to a unitary representation.

In Section 9, we discuss an application of Theorems 2 and 3 to the
Heisenberg group, a very special case of non-commutativity.

In the following we will be concerned mainly with Theorems 2 and
3.

\section{Applications}

All the examples that follows are applications of Theorem 2 and
refer to operators which are not \emph{normal}. This is so because
of the following corollary to Nagy's Theorem 2:

\begin{corollary} A normal operator $T$ either is already unitary
or it is not similar to any unitary operator. Equivalently: A
normal operator $T$ is unitary if and only if it satisfies the
Nagy condition.
\end{corollary}

\begin{proof} \ When $T$ is normal the operators appearing in its
polar decomposition, $T=V|T|,$ commute, so that $|T|$ satisfies
Nagy's condition Eq. (\ref{unifbound}) together with $T$. Then
$|T|$ is both similar to a unitary operator and positive
selfadjoint. This implies that the spectrum of $|T|$ reduces to 1
, so that $|T|=\mathbb{I}$ and $T=V$.
\end{proof}

\textbf{Example 1)} As a simple example consider the group of
translation on the line realized on $L_{2}(\mathbb{R})$ with a
measure which is not translationally invariant, i.e.
\begin{equation}
(T_{t}\Psi )(x):=\Psi (x+t) \ ,\ \Psi\in L_{2}(\mathbb{R},\rho
(x)dx), \end{equation}
where $\rho(x)$ is any function $0<\alpha <\rho (x)<\beta <\infty $ and denote by $%
h_{\rho }$ the corresponding scalar product. If the limit $%
\lim\limits_{x\rightarrow -\infty }\rho (x)$ exists , say $%
\lim\limits_{x\rightarrow -\infty }\rho (x)=a,$ then it is trivial to
compute the Banach limit because it agrees with a limit in the usual sense.
In fact by Lebesgue Theorem we have:
\begin{equation}
\begin{array}{l}
\lim\limits_{t\rightarrow \infty }\int\limits_{\mathbb{R}}\Psi ^{\ast
}(x+t)\Phi (x+t)\rho (x)dx=\lim\limits_{t\rightarrow \infty }\int\limits_{%
\mathbb{R}}\Psi ^{\ast }(x)\Phi (x)\rho (x-t)dx= \\
\int\limits_{\mathbb{R}}\lim\limits_{t\rightarrow \infty }\Psi
^{\ast }(x)\Phi (x)\rho (x-t)dx=a\int\limits_{\mathbb{R}}\Psi
^{\ast }(x)\Phi (x)dx=ah_{0}(\Psi ,\Phi ).
\end{array}
\end{equation}
This shows that the Banach limit gives $h_{T}(\Psi ,\Phi )=ah_{0}(\Psi ,\Phi
),$ i.e. \ it is a multiple of the standard translation invariant scalar
product. Therefore
\begin{equation}
h_{T}(\Psi ,\Phi )=h_{\rho }(Q^{2}\Psi ,\Phi )=h_{\rho }((\sqrt{\frac{a}{%
\rho }})^{2}\Psi ,\Phi )\ \ \ ,
\end{equation}
that is $Q=\sqrt{\frac{a}{\rho }}$ and
\begin{equation}
(U_{t}\Phi )(x)=(QT_{t}Q^{-1}\Phi )(x)=\sqrt{\frac{\rho (x+t)}{\rho (x)}}%
\Phi (x+t)
\end{equation}
is unitary in $L_{2}(\mathbb{R},\rho (x)dx).$

\textbf{Example 2)} The following example deals with a non-diagonalizable uniformly bounded
 operator $T$ defined on $L_{2}(\mathbb{R%
},dx)$ as
\begin{equation}
(T\Psi )(x):=f(x)\Psi (-x)
\end{equation}
where $f(x)$ is a bounded function:
\begin{equation}
0<\alpha \leq |f(x)|\leq \beta <\infty \;\;.
\end{equation}
In other words $T$ is the product of the parity operator $P$ times
the multiplicative bounded operator $f$. Note that $T$ is
non-normal as
\begin{equation}
fPPf^{\ast }-Pf^{\ast }fP=|f(x)|^{2}-|f(-x)|^{2}
\end{equation}
which is not zero for a generic $f$. Moreover this function $f$
has to be chosen in such a way that $T$ satisfies Nagy's
condition. For this, taking into account that:

for $n>0$%
\begin{equation}
(T^{n}\Psi )(x)=\left( \prod\limits_{k=1}^{n}f((-1)^{k+1}x)\right) \Psi
((-1)^{n}x)\ \ \ \ \ \ n>0
\end{equation}

while for $n<0$%
\begin{equation}
(T^{n}\Psi )(x)=\left( \prod\limits_{k=1}^{-n}f^{-1}((-1)^{k}x)\right) \Psi
((-1)^{n}x)\ \ \ \ \ \ n<0
\end{equation}
the condition $||T^{n}||<K$ implies the functional relation $|f(x)f(-x)|=1$
, which admits the general solution :
\begin{equation}
f(x)=\frac{\mu (x)}{\mu (-x)}e^{i\varphi (x)}
\end{equation}
where $\mu (x)$ is a real function such that: $0<\mu _{1}\leq \mu (x)\leq
\mu _{2}<\infty .$

The Banach limit is readily evaluated and results
\begin{equation}
\begin{array}{l}
h_{T}(\Phi ,\Psi )=Lim_{n\rightarrow \infty }h_{0}(T^{n}\Phi ,T^{n}\Psi )=
\\
=\frac{1}{2}\int \Phi ^{\ast }(x)\Psi (x)(1+\frac{\mu ^{2}(-x)}{\mu ^{2}(x)}%
)dx=h_{0}(Q^2\Phi ,\Psi )
\end{array}
\end{equation}
where $Q^2=\frac{1}{2}(1+\frac{\mu ^{2}(-x)}{\mu ^{2}(x)})$ is a
bounded positive operator, as expected. The operator $U_{T}$,
similar to $T$, which is unitary with respect to the standard
scalar product $h_{0}$ is
\begin{equation}
U_{T}=QTQ^{-1}=e^{i\varphi (x)}P\;\;.
\end{equation}
$U_{T}$ has only continuous spectrum given by
\begin{equation}
\lambda _{\pm }=\pm e^{\frac{i}{2}(\varphi (x_{0})+\varphi
(-x_{0}))}\;\;;\;\;x_{0}\in \mathbb{R}
\end{equation}
with corresponding generalized eigenfunctions $\Psi _{\pm }(x)$ :
\begin{equation}
\Psi _{\pm }(x)=e^{\frac{i}{2}\varphi (x_{0})}\delta (x-x_{0})\pm e^{\frac{i%
}{2}\varphi (-x_{0})}\delta (x+x_{0})\;\;.
\end{equation}

\textbf{Example 3)} The following example is similar to the
previous one but parity is now replaced by a translation of a
fixed amount $a$:

\begin{equation}
(T_{a}\Psi )(x):=f(x)\Psi (x+a)
\end{equation}
where $f(x)$ is a bounded function
\begin{equation}
0<\alpha \leq |f(x)|\leq \beta <\infty \ \ \ \ .
\end{equation}
Imposing $||T^{n}||<K$ one gets that $f(x)=g(x)e^{i\phi (x)}$ with
$g(x)$ real and positive such that $g(x+a)=g(x)^{-1}$ and $\phi
(x)$ arbitrary and
real. Then as before one gets $Q^2 =\frac{1}{2}(1+g^{2}(x))$ . The spectrum of $%
T_{a}$ is continuous, indeed the equation:
\begin{equation}
(T_{a}\Psi )(x)=g(x)e^{i\phi (x)}\Psi (x+a)=\mu \Psi (x)
\end{equation}
can be solved in the form \ $\Psi (x)=\sqrt{g(x)}e^{i(\lambda x+\chi (x))}$
; then
\begin{equation}
(T_{a}\Psi )(x)=T_{a}\sqrt{g(x)}e^{i(\lambda x+\chi (x))}=\sqrt{g(x)}%
e^{i(\lambda x+\chi (x+a)+\phi (x))}e^{i\lambda a}\;\;.
\end{equation}
Therefore $\chi (x)$ must fulfill the functional relation $\chi (x+a)=\chi
(x)+\phi (x)$ ; this relation determines $\chi $ on the entire line once it
is arbitrarily chosen on $[0,a]$ . The continuous spectrum is then the
entire circle $\mu =e^{i\lambda a}$ .

\section{Dependence of the Invariant Metric on the Choice of the Initial One}

In this Section we analyze to what extent the invariant metric
$h_{T}$ changes by a change of the starting fiducial metric
$h_{0}$ to a
topologically equivalent one $h_{0}^{\prime }$. Any change of $h_{0}$ to $%
h_{0}^{\prime }$ is parameterized by a positive definite Hermitian operator $%
C$ by the relation
\begin{equation}
h_{0}(x,y)=h_{0}^{\prime }(Cx,y)
\end{equation}
and we get by Banach limits two invariant scalar products $h_{T}(x,y)$ and $%
h_{T}^{\prime }(x,y)$, which are related by a similar relation
\begin{equation}
h_{T}(x,y)=h_{T}^{\prime }(Rx,y)\ .
\end{equation}

\begin{proposition}Consider the above $C$ and $R$ \ and define $A$ \
in the following way:
\begin{equation}
Lim_{n\rightarrow \infty }\ h_{T}^{\prime
}(A_{n}x,T^{n}y)=:F(x,y)=h_{T}^{\prime }(Ax,y)
\end{equation}
where
\begin{equation}
A_{n}=[C,T^{n}]\ \ .
\end{equation}
Then $R=C+A$ and $[A,T]=-[C,T]$.
\end{proposition}

\begin{proof} From the definition it is trivial to show that
$\{A_{n}\}$ is a set of uniformly bounded operators; therefore it
makes sense to compute
the bilinear functional $F(x,y)$ and the operator $A$ \ such that $%
F(x,y)=h_{T}^{\prime }(Ax,y)$ is well defined via Riesz Theorem.
Then it requires only algebraic manipulations to show both of the
following results: $R=C+A$ \ and $[A,T]=-[C,T]$.
\end{proof}

This shows also that $[R,T]=0$, as it should, because any operator
connecting two $T-$invariant Hermitian scalar product necessarily
commutes with $T$. We will see this in the next Section.

\section{Alternative Invariant Hermitian Structures}

Starting with the automorphism $T$, we may investigate for the
existence of alternative Hermitian structures invariant under the
$\mathbb{Z}-$group action generated by $T.$ Such a $T$ could than
be said to be a bi-unitary map.

Assume therefore that $T$ is uniformly bounded. For the time being
we assume in addition that $T$ is diagonalizable and multiplicity
free (i.e. the commutant of $T$ is Abelian\cite{Simon}). Choose
then a bounded transformation $S$ with bounded inverse such that
\begin{equation}
T=S^{-1}US\ \ .
\end{equation}
where $U$ is unitary. Theorem 2 guarantees that at least one such
$S$ exists. Then
\begin{equation}
h^{S}(x,y):=h_{0}(S^{\dagger }Sx,y)
\end{equation}
turns $T$ into a unitary operator. Because of the uniformly boundedness of $%
T $ , we may also use the Banach limit
\begin{equation}
Lim_{n\rightarrow \infty }h_{0}\left( T^{n}x,T^{n}y\right) =:h_{T}(x,y)
\end{equation}
to define a new invariant Hermitian structure.

By using an $h_{0}-$orthonormal basis $\left\{ \varphi
_{k}\right\} $ of eigenvectors of $U$
\begin{equation}
U\varphi _{k}=e^{i\lambda _{k}}\varphi _{k}\ \ \ ;\ \ h_{0}(\varphi
_{k},\varphi _{j})=\delta _{k,j}
\end{equation}
we find by explicit computation of the Banach limit:
\begin{equation}
h_{T}(x,y)=\sum\limits_{k}h_{0}(Sx,\varphi _{k})h_{0}(\varphi
_{k},Sy)||S^{-1}\varphi _{k}||^{2} ,  \label{blim}
\end{equation}
while:
\begin{equation}
h^{S}(x,y)=\sum\limits_{k}h_{0}(Sx,\varphi _{k})h_{0}(\varphi
_{k},Sy).
\end{equation}

We see that the invariant Banach scalar product $h_{T}$ is obtained from $%
h^{S}$ scaling each $\varphi _{k}$ by a factor $||S^{-1}\varphi
_{k}||.$ Notice that any bounded sequence of positive numbers can
be used in the same way to scale $\varphi _{k}$ to obtain
alternative invariant scalar products.
Thus the sequence $\{1,1,1\dots \}$ corresponds to $h^{S}(x,y)$ while $%
\{||S^{-1}\varphi _{1}||,||S^{-1}\varphi _{2}||,\dots \}$ to $h_{T}(x,y)$ .

If the eigenvalues of $U$ \ are not multiplicity free we need another index $%
l$ to label eigenvectors; then Eq. (\ref{blim}) becomes

\begin{equation}
h_{T}(x,y)=\sum\limits_{k,l,n}h_{0}(Sx,\varphi _{kl})h_{0}(\varphi
_{kn},Sy)h_{0}(S^{-1}\varphi _{kl},S^{-1}\varphi _{kn}).
\end{equation}
As in the multiplicity free case one obtains invariant scalar products
replacing $h_{0}(S^{-1}\varphi _{kl},S^{-1}\varphi _{kn})$ with any bounded
sequence of positive matrices.

What we learn from this example is that we can look for
alternative Hermitian structures on each eigenspace of $U$ and
then combine them with arbitrary positive coupling coefficients
provided that bounded vectors remain bounded with respect to the
newly defined products ( boundedness of the sequence). In this
respect see our previous paper\cite{Morandi} where similar
considerations came out within the finite dimensional situation.

In particular this procedure shows that we may start with $h_{0}$ already
invariant and construct $S$ out of constants of the motion for $T,$ in this
way we would obtain alternative descriptions whenever the sequence $||$ $%
S^{-1}\varphi _{k}||$ is appropriate. For instance in a central force
problem $2+\sin (J^{2}),$ in a basis for the angular momentum, would give $%
2+\sin j(j+1)$ and $1/[$ $2+\sin j(j+1)]$ \ would be an appropriate sequence.

\textbf{Remark} From what we said, it is clear that instead of a
once-for-all chosen sequence of bounded positive numbers (for
instance like $\{1,1,1\dots \}$ and $\{||S^{-1}\varphi
_{1}||,||S^{-1}\varphi _{2}||,\dots \}$ ) we could use a
''point-dependent'' sequence so that the Hermitian metric we
define will be dependent on the point and therefore the ''energy
function'' we associate with linear transformations will not be
quadratic anymore.\cite{Dubrovin} In this way, vector fields are
still linear and therefore compatible with the dynamical linear
superposition rule for state vectors, however the associated
Hamiltonian functions (infinitesimal generators) are no more
homogeneous of degree two. One is getting a kind of
''non-linearity'', one should compare this situation with the one
proposed by Weinberg.\cite {Weinberg}

Having learned from the diagonalizable case, we can obtain a class
of invariant scalar products dropping this assumption. Suppose
again $T$ uniformly bounded. Then
\begin{equation}
T=Q^{-1}UQ
\end{equation}
and
\begin{equation}
h_{T}(x,y)=h_{0}(Q^{2}x,y)\ \ .
\end{equation}
Consider the spectral decomposition of $U$ :
\begin{equation}
U=\int\limits_{0}^{2\pi }e^{i\lambda }dE_{\lambda }^{U}\ \ \ .
\end{equation}
Choose now any positive bounded function $\varphi $ on $[0,2\pi ]$ and
define, in analogy with the diagonalizable case, the scalar product:
\begin{equation}
h_{\varphi }(x,y)=\int\limits_{0}^{2\pi }\varphi (\lambda
)h_{0}(Qx,dE_{\lambda }^{U}Qy)
\end{equation}
One checks easily that $h_{\varphi }(x,y)$ is also invariant and that $%
\varphi =1$ corresponds to the Nagy product $h_{T}(x,y)$ .

Via Riesz Theorem, we may write $h_{\varphi
}(x,y)=h_{T}(C_{\varphi }x,y)$ \ and solve for $C_{\varphi }$. To
this aim we define $B$ as
\begin{equation}
B=\int\limits_{0}^{2\pi }\varphi (\lambda )dE_{\lambda }^{U}
\end{equation}
to get
\begin{equation}
\begin{array}{l}
h_{\varphi }(x,y)=\int\limits_{0}^{2\pi }\varphi (\lambda
)h_{0}(Qx,dE_{\lambda }^{U}Qy)=h_{0}(Qx,BQy)= \\
=h_{0}(Qx,QQ^{-1}BQy)=h_{T}(x,Q^{-1}BQy).
\end{array}
\end{equation}
This formula furnishes $C_{\varphi }=Q^{-1}BQ$ .

It is not hard to show that $T$ commutes with $C_{\varphi }$ , as

\begin{equation}
h_{\varphi }(Tx,Ty)=h_{\varphi }(x,y)=h_{T}(C_{\varphi
}x,y)=h_{T}(C_{\varphi }Tx,Ty)=h_{T}(TC_{\varphi }x,Ty)
\end{equation}
so that $[T,C_{\varphi }]=0$ follows.

We conclude this Section by noting that the class of invariant
scalar product compatible with the starting one is parameterized
by the definite positive elements in the commutant of $T$, and in
particular it is empty only when $T$ is not uniformly bounded$.$

\section{Invariant Hermitian Structures for Commuting Uniformly Bounded
Operators}

The method of finding an invariant scalar product for an automorphism $T$
via a limiting procedure ''at infinity'' applies also in the case of many
uniformly bounded commuting automorphisms.

We analyze first the case of two uniformly bounded commuting
operators $T_{1} $ , $T_{2}$ . They generate an action of the
Abelian group $\mathbb{Z}\times \mathbb{Z}$ by uniformly bounded
operators; Theorem 5 guarantees the existence of an invariant
Hermitian scalar product. We show how to compute one of them.

One first compute $h_{T_{1}}$ as a Banach limit; this is by
construction invariant under $T_{1}$ \ but in general not under
$T_{2}$ . As second step one defines:
\begin{equation}
h_{12}(x,y)=Lim_{n\rightarrow \infty }\ h_{T_{1}}(T_{2}^{n}x,T_{2}^{n}y)\ .
\label{h12}
\end{equation}

\begin{proposition} Consider two uniformly bounded commuting
operators $T_{1}$ , $T_{2}$. Then
\begin{description}
\item[(i)] $h_{1,2}$, defined in Eq. (\ref{h12}), is
invariant under the action of the entire $\ \mathbb{Z}\times \mathbb{Z}$ .

\item[(ii)]  If $T_{1}$ is multiplicity free, $h_{T_{1}}$ is already
invariant under the action of the entire $\ \mathbb{Z}\times \mathbb{Z}$.
\end{description}
\end{proposition}

\begin{proof} The proof of (i) is trivial. To prove (ii) consider
$C_{1}$ defined through
\begin{equation}
h_{T_{1}}(x,y)=:h_{12}(C_{1}x,y)\ .  \label{Cformula}
\end{equation}
As shown before $C_{1}$ commutes with $T_{1}$ , but $T_{1}$ is multiplicity
free and therefore its commutant is Abelian. So $T_{2}$ must commute with $%
C_{1}$ and the result follows at once from Eq.
(\ref{Cformula}).
\end{proof}

\textbf{Remark: }These results may obviously be extended to any uniformly
bounded action of the product of a finite number of groups $\mathbb{Z}$
and/or $\mathbb{R}$ .

\begin{corollary} Suppose that, in a uniformly bounded action of
an Abelian group, a particular operator $\widetilde{T}$ is
multiplicity free. Then the corresponding Nagy product
$h_{\widetilde{T}}$ \ is invariant for the entire action.
\end{corollary}

\begin{proof}The result follows readily extending the argument
of the proof of the above point (ii) to the general case of
Theorem 5.
\end{proof}

\section{Relating Two Uniformly Bounded Operators via Banach Limits}

The uniform boundedness condition is necessary and sufficient for
an operator
$T$ to be similar to a unitary one $U$ as stated in Theorem 2, then $T$ and $%
U$ have the same spectrum. The $A-$relatedness property is a
condition weaker than similarity that has been utilized to discuss
in more general terms operators and relations between their
spectra, see for instance Ref. 17. We recall that two operators
 $T_{1}$ and $T_{2}$ are said to be $%
A- $related if
\begin{equation}
T_{1}A=AT_{2};
\end{equation}
then the operator $A$ is called an intertwining operator.

In this Section we discuss the case of two not necessarily
commuting operators $T_{1}$ and $T_{2}$ both uniformly bounded and
ask for conditions under which there exists a nontrivial operator
$A$ intertwining them. We require also that $A$ should be bounded.

For this discussion consider the bilinear functional $\mathcal{F}(x,y)$
defined as follows:
\begin{equation}
\mathcal{F}(x,y):=Lim_{n\rightarrow \infty }h_{0}(T_{2}^{n}x,T_{1}^{n}y)
\end{equation}
this limit is well defined for all $x,y\in \mathbb{H}$ \ and a
simple computation shows that $|\mathcal{F}(x,y)|\leq K||x||\
||y||$ .

Using Riesz Theorem we define three bounded operators $A_{0}$, $A_{1}$, $%
A_{2}$ via:
\begin{equation}
\mathcal{F}(x,y)=h_{0}(A_{0}x,y)=h_{T_{1}}(A_{1}x,y)=h_{T_{2}}(A_{2}x,y).
\end{equation}
The following proposition holds:

\begin{proposition} With $\mathcal{F}$, $A_{0},$ $A_{1},$ $A_{2}$
defined as above the relations 1-4 hold:

\begin{enumerate}
\item  $A_{0}=Q_{1}^{2}A_{1}=Q_{2}^{2}A_{2}$ ;

\item  $A_{0}T_{2}=(T_{1}^{\dagger })^{-1}A_{0}$ \ , the adjoint is with
respect to the $h_{0}$ scalar product;

\item  $A_{1}T_{2}=(T_{1}^{\dagger })^{-1}A_{1}$ \ , the adjoint is with
respect to the $h_{T_{1}}$ scalar product;

\item  $A_{2}T_{2}=(T_{1}^{\dagger })^{-1}A_{2}$ \ , the adjoint is with
respect to the $h_{T_{2}}$ scalar product.
\end{enumerate}
\end{proposition}

\begin{proof} These relations follow at once from the definitions
of the $A$'s operators and the invariance property of $\mathcal{F}$ : $\mathcal{F}%
(T_{2}x,T_{1}y)=\mathcal{F}(x,y)$.
\end{proof}

\textbf{Remark.} Relation (3) shows that $T_{1}$ and $T_{2}$ are
$A_{1}-$related when $\mathcal{F\neq }0$, because in this case it results that $%
(T_{1}^{\dagger })^{-1}=T_{1}$.

We therefore are led to examine some conditions which guarantees that $%
\mathcal{F\neq }0.$ Consider the simple case of $T_{1}$ and $T_{2}$ both
diagonalizable and multiplicity free. Recalling that in our hypothesis on $%
T_{1}$ and $T_{2}$ we have:
\begin{equation}
T_{1}=Q_{1}^{-1}U_{1}Q_{1}\ \ \ ,\ \ T_{2}=Q_{2}^{-1}U_{2}Q_{2}\ \ ,
\end{equation}
it is easy, by using two suitable orthonormal basis, to obtain the following
expression for $\mathcal{F}$ :
\begin{equation}
\mathcal{F}(x,y)=\sum\limits_{k,q}\delta _{\lambda _{k},\mu
_{q}}h_{0}(x,Q_{2}\psi _{q})\ h_{0}(Q_{2}^{-1}\psi _{q},Q_{1}^{-1}\varphi
_{k})\ h_{0}(Q_{1}\varphi _{k},y)  \label{Flimit}
\end{equation}
where $U_{1}\varphi _{k}=\lambda _{k}\varphi _{k}$\ , $U_{2}\psi _{q}=\mu
_{q}\psi _{q}$ \ and $h_{0}(\varphi _{k},\varphi _{j})=\delta _{k,j}$ , $%
h_{0}(\psi _{k},\psi _{j})=\delta _{k,j}$ .

We have shown therefore that

\begin{proposition} If $T_{1}$ and $T_{2}$ are uniformly bounded,
diagonalizable and multiplicity free operators then
$\mathcal{F\neq }0$ if and only if they have at least one common
eigenvalue and $h_{0}(Q_{2}^{-1}\psi _{q},Q_{1}^{-1}\varphi
_{k})\neq 0$.
\end{proposition}

We note also that from Eq. (\ref{Flimit}) it is easy to obtain an
explicit expression for the intertwining operators $A$'s. For
instance:
\begin{equation}
A_{0}=\sum\limits_{k,q}\delta _{\lambda _{k},\mu _{q}}\
h_{0}(Q_{1}^{-1}\varphi _{k},Q_{2}^{-1}\psi _{q})h_{0}(Q_{2}\psi _{q},\cdot
)Q_{1}\varphi _{k}\ \ \ .  \label{interwining}
\end{equation}
An argument similar to the one used after Eq. (\ref{blim}) shows
that one can replace in Eq. (\ref{interwining})
$h_{0}(Q_{1}^{-1}\varphi _{k},Q_{2}^{-1}\psi _{q})$ with any
bounded sequence of numbers, obtaining in this way other
intertwining operators.

One can use Eq. (\ref{interwining}) even in cases involving
continuous spectra. For instance in $L_{2}(\mathbb{R},dx)$
consider $(T_{1}\Psi )(x)=e^{ix}\Psi (x)$ and \ $(T_{2}\Psi
)(x)=\Psi (x+a)$ \ , $a\in \mathbb{R}$; then, when the sequence
$h(Q_{1}^{-1}\varphi _{k},Q_{2}^{-1}\psi _{q})$ is replaced by a
constant sequence, Eq. (\ref{interwining}) leads to the Fourier
transform operator.

\section{Invariant Hermitian Structures for Realizations of the Heisenberg
Group}

Elsewhere we have shown\cite{Romp2001} how alternative symplectic
structures, on a finite dimensional real symplectic\ vector space
$V$ , give rise to alternative Weyl systems, i.e. alternative
projective unitary representations of the Abelian vector group
$V.$

In this Section we show that it is possible to find invariant
Hermitian structures for any realization of the Heisenberg group
in terms of uniformly bounded operators.

We consider operators $T_{1}$ , $T_{2}$ , $T_{3}$  uniformly
bounded and obeying the following commutation relations
\begin{equation}
T_{1}T_{3}T_{1}^{-1}T_{3}^{-1}=\mathbb{I}=T_{2}T_{3}T_{2}^{-1}T_{3}^{-1}
\end{equation}
and
\begin{equation}
T_{1}T_{2}T_{1}^{-1}T_{2}^{-1}=T_{3}\ \ .
\end{equation}

\begin{proposition} For $T_{1}$ , $T_{2}$ , $T_{3}$ satisfying the
stated conditions it is possible to find a Hermitian structure
that converts them into unitary operators.
\end{proposition}

\begin{proof} There is a scalar product, say $\ h_{13}$, that makes $T_{1}$ and
$T_{3}$ unitary since they commute and are uniformly bounded; in
this scalar product $T_{2}$ will not be unitary in general.
Consider therefore
\begin{equation}
h(x,y):=Lim_{n\rightarrow \infty }h_{13}(T_{2}^{n}x,T_{2}^{n}y).
\end{equation}
This scalar product $h$ makes $T_{2}$ unitary and one checks
easily that it leaves unitary $T_{1}$ and $T_{3}$ as well. In fact
\begin{eqnarray}
h(T_{1}x,T_{1}y) &=&Lim_{n\rightarrow \infty
}h_{13}(T_{2}^{n}T_{1}x,T_{2}^{n}T_{1}y)=  \nonumber \\
&=&Lim_{n\rightarrow \infty
}h_{13}(T_{1}T_{3}^{n}T_{2}^{n}x,T_{1}T_{3}^{n}T_{2}^{n}y)=h(x,y)
\end{eqnarray}
and the same for $T_{3}$ .
\end{proof}

In the future we will show the use of this proposition, to compare
alternative Weyl systems when they are considered on the same
Hilbert space.

\section{Quantum Systems as Hamiltonian Systems}

Let us recall that for a complex Hilbert space $\mathbb{H}$ with
Hermitian structure $h$ it is possible to define a symplectic
structure by setting
\begin{equation}
\omega _{h}(x,y):=\frak{Im}\ \ h(x,y)\ \ \ .
\end{equation}

Given a symplectic vector space $(V,\omega )$ we may define a
Poisson Bracket on $V^{\ast }$ by defining it first on linear
functions and then by using the Leibnitz rule on all
differentiable functions $\mathcal{F}(V^{\ast })$ . On
$Lin(V^{\ast },\mathbb{C})\subset \mathcal{F}(V^{\ast })$ we set
\begin{equation}
\left\{ v_{1},v_{2}\right\} =\omega (v_{1},v_{2})
\end{equation}
where on the left hand side $v_{1},v_{2}\in \mathcal{F}(V^{\ast })$ and on
the right $v_{1},v_{2}\in V$ .

More directly, on any totally reflexive space\cite{Nelson}, it is
possible to define a non-degenerate Poisson Bracket if the space
is strongly symplectic.\cite{Abraham} Indeed introducing
differentials of functions
\begin{equation}
df:V^{\ast }\rightarrow V^{\ast }\times V^{\ast \ast }\equiv
V^{\ast }\times V
\end{equation}
in intrinsic form at each point $\alpha \in V^{\ast },$ we define
\begin{equation}
\left\{ f,g\right\} (\alpha ):=\omega (df(\alpha ),dg(\alpha ))\ \ .
\end{equation}

In our setting $V$ is a Hilbert space and therefore there is a non intrinsic
isomorphism between $V$ and $V^{\ast }$ so that we have a Poisson Bracket
defined also on $\mathcal{F}(V).$ It follows easily that a complex unitary
linear transformation $U$ on $\mathbb{H}$ is symplectic. A densely defined
complex linear operator $\Gamma $ is a Hamiltonian vector field iff $\ H$ is
Hermitian. The Hamiltonian function associated with $\Gamma $ is given by
the formula
\begin{equation}
f_{H}(\psi )=\frac{1}{2}h(H\psi ,\psi )\ \ \ \ \psi \in D(\Gamma )\ \ .
\end{equation}
Thus, we associate an Hamiltonian function (infinitesimal
generating function) with any vector field which preserves the
Hermitian structure $h$.

Disregarding domain problems, for the moment, it is easy to show that if $%
A,B,.....$ are Hermitian operators, the associated Hamiltonian functions,
say $f_{A},f_{B},....,$ satisfy the Poisson Bracket relations
\begin{equation}
\left\{ f_{A},f_{B}\right\} (\psi )=f_{i[A,B]}(\psi )\ \ \ .
\end{equation}
It is now possible to investigate the consequence of the existence
of alternative invariant Hermitian structures. With the notation
of Theorem 2, we have a new Poisson Bracket defined by
\begin{equation}
\left\{ v_{1},v_{2}\right\} _{T}=i\omega _{h}(Q^{2}v_{1},v_{2})\ \ \ .
\end{equation}
These new metrics will associate alternative quadratic functions
with every operator $A.$ All of the induced Poisson Bracket on
quadratic functions will be pairwise compatible in the sense of
bi-Hamiltonian systems. In the Ehrenfest description of quantum
dynamics, we have
\begin{equation}
i\hbar \frac{d}{dt}f_{A}=\left\{ f_{H},f_{A}\right\}
\end{equation}
and therefore the same vector field will be given different Hamiltonian
descriptions, with $f_{H}$ \ and the Poisson Bracket depending on the chosen
invariant metric.

The relation $h(v_{1},v_{2})=h^{^{\prime }}(Rv_{1},v_{2})$,
between two invariant Hermitian forms, suggests the correspondence
between operators
\begin{equation}
N:A\mapsto RA
\end{equation}

According to the Ref. 32, we may define a new associative product
on the space of operators
\begin{equation}
A\circ _{N}B=N(A)B+AN(B)-N(AB)
\end{equation}
which gives $A\circ _{N}B=ARB.$ Any time that $R$ is a constant of the
motion for $H,$ as in this case, we obtain a new alternative associative
product on the space of operators which makes $H$ into an inner derivation.

In conclusion, we have shown how alternative Hamiltonian descriptions for
the Schroedinger equation give rise to alternative description in the
Ehrenfest picture and the Heisenberg picture.

\section{Conclusions}

In this paper we have addressed the problem of alternative quantum
Hamiltonian descriptions of the same vector field on the space of
quantum states. In this way we have avoided dealing with the
ambiguity of quantization procedures for classical bi-Hamiltonian
systems.

We have briefly addressed the question of the alternative
descriptions at the level of the Ehrenfest picture and Heisenberg
picture. By using this results, in the future we shall consider
the quantum-classical transition to show how these alternative
description at the quantum level will reproduce known alternative
Hamiltonian descriptions for the corresponding classical systems.

\end{document}